%% file: main.tex
\documentclass[runningheads]{llncs}

\usepackage{hyperref}
\usepackage{xcolor}
\usepackage{graphicx}
\usepackage{eso-pic}
\usepackage{algorithm}
\usepackage[noend]{algpseudocode}
\usepackage{array, makecell} 
\usepackage{multirow}
\usepackage{epstopdf}
\usepackage{ulem}
\usepackage{caption}
\usepackage{footnote}
\usepackage{booktabs}
\usepackage{amsmath}
\usepackage{amsfonts}
\usepackage{amssymb}
\usepackage{tikz}
\usetikzlibrary{shapes,automata, positioning, arrows}

\usepackage{mathrsfs, mathtools}

\newcommand{\ignore}[1]{}

\title{An AMM minimizing user-level extractable value and loss-versus-rebalancing}

\usepackage{authblk}
\author{Conor McMenamin\inst{1} \and Vanesa Daza\inst{1,2}}

\institute{Department of Information and Communication Technologies, Universitat Pompeu Fabra, Barcelona, Spain \and
CYBERCAT - Center for Cybersecurity Research of Catalonia}

\authorrunning{McMenamin and Daza}
\titlerunning{V0LVER}

\newcommand\nnfootnote[1]{%
  \begin{NoHyper}
  \renewcommand\thefootnote{}\footnote{#1}%
  \addtocounter{footnote}{-1}%
  \end{NoHyper}
}

\input{Variables}

\begin{document}

\maketitle

\nnfootnote{\begin{minipage}{0.06\textwidth}
    \includegraphics[width=\linewidth]{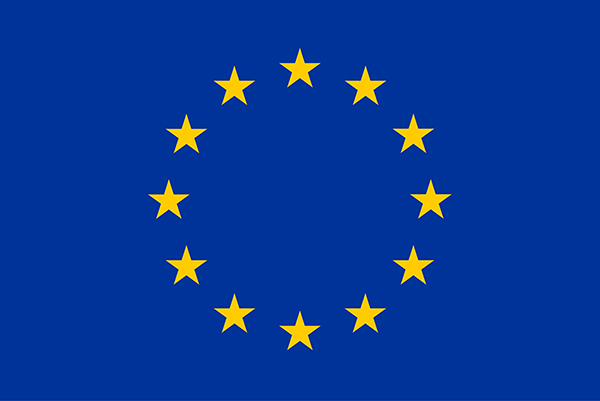}
    \end{minipage}%
    \hfill%
    \begin{minipage}{0.88\textwidth} This Technical Report is part of a project that has received funding from the European Union's Horizon 2020 research and innovation programme under grant agreement number 814284 \end{minipage}}

\input{Abstract}


\input{Introduction/Introduction}

\input{Appendices/RelatedWork}

\input{Prelims/Prelims}
\input{FormalProtocol}
\input{ProtocolProperites}
\input{Discussion}

\input{Conclusion}

\addcontentsline{toc}{section}{Bibliography}
\bibliographystyle{splncs04}
\bibliography{references}

\end{document}

%% file: Variables.tex
\usepackage{cancel}

\usepackage{pifont}

\def\bitcoin{%
  \leavevmode
  \vtop{\offinterlineskip 
    \setbox0=\hbox{B}%
    \setbox2=\hbox to\wd0{\hfil\hskip-.03em
    \vrule height .3ex width .15ex\hskip .08em
    \vrule height .3ex width .15ex\hfil}
    \vbox{\copy2\box0}\box2}}

\newcommand{\height}{H}

\newcommand{\MIFP}{\epsilon}

\newcommand{\player}{\textit{P}}

\newcommand{\protocolName}{\text{V0LVER}}

\newcommand{\revealTXDelay}{T}

%% file: Abstract.tex
\begin{abstract}
  We present $\protocolName$, an AMM protocol which solves an incentivization trilemma between users, passive liquidity providers, and block producers. $\protocolName$ enables users and passive liquidity providers to interact without paying MEV or incurring uncontrolled loss-versus-rebalancing to the block producer. $\protocolName$ is an AMM protocol built on an encrypted transaction mempool, where transactions are decrypted after being allocated liquidity by the AMM. $\protocolName$ ensures this liquidity, given some external market price, is provided at that price in expectancy. This is done by incentivizing the block producer to move the pool price to the external market price. With this, users transact in expectancy at the external market price in exchange for a fee, with AMMs providing liquidity in expectancy at the external market price. Under block producer and liquidity provider competition, all of the fees in $\protocolName$ approach zero. Without block producer arbitrage, $\protocolName$ guarantees fall back to those of an AMM, albeit free from loss-versus-rebalancing and user-level MEV.  
\end{abstract}

%% file: Introduction/Introduction.tex
\section{Introduction}

AMMs have emerged as a dominant medium for decentralized token exchange. This is due to several important properties making them ideal for decentralized liquidity provision. AMMs are efficient computationally, have minimal storage needs, matching computations can be done quickly, and liquidity providers (LPs) can be passive. Thus, AMMs are uniquely suited to the severely computation- and storage-constrained environment of blockchains. 

Unfortunately, the benefits of AMMs are not without significant costs. For users sending orders to an AMM, these orders can be front-run, sandwiched, back-run, or censored by the block producer in a phenomenon popularized as MEV \cite{FlashBoys2.0}. Current estimates for MEV against AMM users on Ethereum are upwards of $\$600$M \cite{QuantifyingExtractableValueGervais,FlashbotsExploreWebsite}. By the nature of AMMs and their continuous liquidity curves, the amount of MEV extractable from an order is increasing in order impact (related in large part to order size and slippage tolerance). Thus, MEV effectively caps the trade size allowable on current AMMs when compared to the costs for execution on MEV-protected centralized exchanges. This is a critical barrier for DeFi, and blockchain adoption in general. 

\begin{figure}
    \centering
    \includegraphics[scale=0.45]{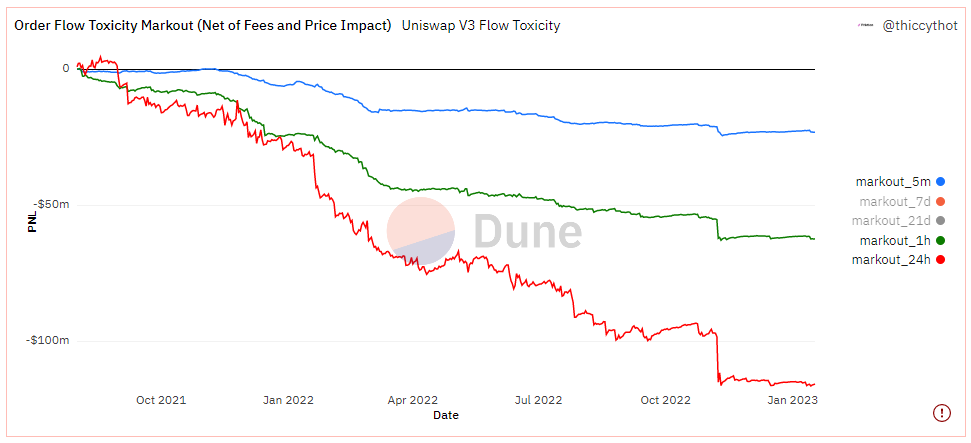}
    \caption[Toxicity of Uniswap V3 Order Flow \cite{duneToxicityQuery}.]{Toxicity of Uniswap V3 Order Flow \cite{duneToxicityQuery}. This graph aggregates the PnL of all trades on the Uniswap V3 WETH/USDC pool, measuring PnL of each order after 5 minutes, 1 hour, and 1 day. These are typical time periods within which arbitrageurs close their positions against external markets. This demonstrates the current losses being suffered by AMM pools are significant, consistent, and unsustainable. As LVR is significant and consistent, a large part of these losses can be prevented by minimizing LVR. }
    \label{fig:Toxic}
\end{figure}

Another one of these significant costs for AMMs is definitively formalized in \cite{LVRRoughgarden} as \textit{loss-versus-rebalancing} (LVR). It is proved that as the underlying price of a swap moves around in real-time, the discrete-time progression of AMMs leave arbitrage opportunities against the AMM. In centralized finance, market makers (MMs) typically adjust to new price information before trading. This comes at a considerable cost to AMMs (for constant function MMs (CFMMs), \cite{LVRRoughgarden} derives the cost to be quadratic in realized moves), with similar costs for AMMs derived quantitatively in \cite{Park2021TheCF,DEXCritiqueCapponi}. These costs are being realized by LPs in current AMM protocols. 
Furthermore, toxic order flow, of which LVR is a prime example, is consistently profiting against AMM LPs (Figure \ref{fig:Toxic}).

These costs are dooming DeFi, with current AMM design clearly unsatisfactory. In this paper, we provide V0LVER, an AMM protocol which formally protects against both MEV and LVR. beckoning a new era for AMMs, and DeFi as a whole.

\input{Introduction/OurContribution}

%% file: Introduction/OurContribution.tex
\subsection{Our Contribution}

In this paper we introduce $\protocolName$ \footnote{near-\textbf{0} \textbf{E}xtractable \textbf{V}alue and \textbf{L}oss-\textbf{V}ersus-\textbf{R}ebalancing $\leadsto$ \textbf{V0LVER }}, an AMM which provides arbitrarily high protection against user-level MEV and LVR. $\protocolName$ is the first AMM to align the incentives of the three, typically competing, entities in AMMs; the user, the pool, and the block producer. 
This is done by ensuring that at all times, a block producer is incentivized to move the pool to the price maximizing LVR. When the block producer chooses a price, the block producer is forced to assert this is correct, a technique introduced in \cite{DiamondMcMenamin}. Unfortunately, the protocol in \cite{DiamondMcMenamin} gives the block producer total power to extract value from users, due to order information being revealed to the block producer before it is allocated a trading price in the blockchain. To address this, $\protocolName$ is built on an encrypted mempool. Modern cryptographic tools allow us to encrypt the mempool using zero-knowledge based collateralized commit-reveal protocols \cite{DPACCsMcMenamin,ZCash,FairTraDEXMcMenamin,TornadoCashWebsite}, delay encryption \cite{DelayEncryptionBurdges,FairPoSChiang} and/or threshold encryption \cite{HelixAsayag}. We assume the existence of such a mempool within which all sensitive order information is hidden until the order has been committed a price against the AMM. Given these encrypted orders, we demonstrate that a block producer forced to show liquidity to such an order maximizes her own utility by showing liquidity centred around the external market price (bid below the price and offered above the price).\footnote{This holds true in many CFMMs, including the famous Uniswap V2 protocol \cite{Adams2020UniswapVC}} 

As such, the external market price is the price point maximizing the block producers LVR extraction (due to the replicated LVR protection of \cite{DiamondMcMenamin}), around which  profit is maximized when forced to trade against some (varying) percentage of indistinguishable orders. This strictly incentivizes block producers to move the price of a $\protocolName$ pool to the external market price. This provides users with an AMM where the expected trade price in the presence of arbitrageurs is always the external market price, excluding fees, and the LVR against the pool is minimized when these arbitrageurs are competing. Although batching orders against AMM liquidity has been proposed as a defense against LVR \cite{BatchCFMMsRamseyer}, naively batching orders against an AMM still allows a block producer to extract LVR by censoring user orders. In $\protocolName$, block producers are effectively forced to immediately repay LVR, while being incentivized to include order commitments in the blockchain and allocate liquidity to these orders through the AMM.

%% file: Appendices/RelatedWork.tex
\section{Related Work}\label{sec:RW}

As the phenomenon of LVR has only recently been identified, there are only two academic papers on the subject of LVR protection \cite{DynamicAMMsKrishnamachari,DiamondMcMenamin} to the best of our knowledge, with no work protecting against both LVR and user-level MEV.

In \cite{DynamicAMMsKrishnamachari}, the AMM must receive the price of a swap from a trusted oracle before users can interact with the pool. Such sub-block time price data requires centralized sources which are prone to manipulation, or require the active participation of AMM representatives, a contradiction of the passive nature of AMMs and their liquidity providers. We see this as an unsatisfactory dependency for DeFi protocols.

Our work is based on some of the techniques of the Diamond protocol as introduced in \cite{DiamondMcMenamin}. The Diamond protocol requires block producers to effectively attest to the final price of the block given the orders that are to be proposed to the AMM within the block. This technique requires the block producer to know exactly what orders are going to be added to the blockchain. This unfortunately gives the block producer total freedom to extract value from users submitting orders to the AMM. With $\protocolName$, we address this issue while keeping the LVR protection guarantees of Diamond.

Encrypting the transaction mempool using threshold encryption controlled by a committee has been proposed in \cite{HelixAsayag} and applied in \cite{Penumbra}. In \cite{Penumbra}, a DEX involving an AMM and based on frequent batch auctions \cite{FrequentBatchAuctionsBudish} is proposed. This DEX does not provide LVR resistance, and incentivizes transaction censorship when a large LVR opportunity arises on the DEX. This is protected against in $\protocolName$.

%% file: Prelims/Prelims.tex
\section{Preliminaries}

This section introduces the key terminology and definitions needed to understand LVR,  and the proceeding analysis. In this work we are concerned with a single swap between token $x$ and token $y$. We use $x$ and $y$ subscripts when referring to quantities of the respective tokens. The external market price of a swap is denoted by $\MIFP$, with the price of a swap quoted as the quantity of token $x$ per token $y$. 

\subsection{Constant Function Market Makers}\label{sec:CFMMs}

A CFMM is characterized by \textit{reserves} $(R_x,R_y) \in \mathbb{R}_{+}^2$ which describes the total amount of each token in the pool. The price of the pool is given by \textit{pool price function} $P: \mathbb{R}^2_{+} \rightarrow \mathbb{R}$ taking as input pool reserves $(R_x,R_y)$. $P()$ has the following properties:
\begin{equation}\label{eq:pricerestrictions}
    \begin{aligned}
        \text{(a)}&\  P() \text{ is everywhere differentiable, with }\frac{\partial P}{\partial R_x}>0,\ \frac{\partial P}{\partial R_y}<0. \\
        \text{(b)}& \ \lim_{R_x \rightarrow 0} P = 0, \ \lim_{R_x \rightarrow \infty} P = \infty, \ \lim_{R_y \rightarrow 0} P = \infty, \ \lim_{R_y \rightarrow \infty} P = 0. \\
        \text{(c)}& \ \text{If } P(R_x,R_y)=p, \text{ then } P(R_x+ cp, R_y + c )=p, \ \forall c >0. \\
    \end{aligned}
\end{equation}

For a CFMM, the \textit{feasible set of reserves} $C$ is described by:
\begin{equation}
    C = \{ (R_x,R_y) \in \mathbb{R}_{+}^2 : f(R_x,R_y)=k \}
\end{equation}
where $f:\mathbb{R}_{+}^2 \rightarrow \mathbb{R}$ is the pool invariant and $k \in \mathbb{R}$ is a constant. The pool is defined by a smart contract which allows any player to move the pool reserves from the current reserves $(R_{x,0},R_{y,0}) \in C$ to any other reserves $(R_{x,1},R_{y,1}) \in C$ if and only if the player provides the difference $(R_{x,1}- R_{x,0} ,R_{y ,1}-R_{y,0})$.

Whenever an arbitrageur interacts with an AMM pool, say at time $t$ with reserves $(R_{x,t},R_{y,t})$, we assume as in \cite{LVRRoughgarden} that the arbitrageur always moves the pool reserves to a point which maximizes arbitrageur profits, exploiting the difference between $P(R_{x,t},R_{y,t})$ and the external market price at time $t$, denoted $\MIFP_t$. Therefore, the LVR between two blocks $B_t$ and $B_{t+1}$ where the reserves of the AMM at the end of $B_t$ are $(R_{x,t},R_{y,t})$ and the external market price when creating block $B_{t+1}$ is $\MIFP_{t+1}$ is:
\begin{equation}\label{eq:LVR}
    R_{x,t}-R_{x,t+1}+ (R_{y,t}-R_{y,t+1})\MIFP_{t+1}.
\end{equation}
In this paper, we consider only the subset of CFMMs in which, given the LVR extracted in block $B_{t+1}$ corresponds to reserves $(R_{x,t+1},R_{y,t+1})$, $P(R_{x,t+1},R_{y,t+1})$ $=\MIFP_{t+1}$. This holds for Uniswap V2 pools, among others.

\subsection{LVR-resistant AMM}\label{sec:Diamond}

We provide here an overview of the most important features of Diamond \cite{DiamondMcMenamin}, an AMM protocol which is proved to provide arbitrarily high LVR protection under competition to capture LVR among block producers. In $\protocolName$, we adapt these features for use on an encrypted transaction mempool.

A Diamond pool $\Phi$ is described by reserves $(R_{x},R_{y})$, a pricing function $P()$, a pool invariant function $f()$, an \textit{LVR-rebate parameter} $\beta \in (0,1)$, and \textit{conversion frequency} $T \in \mathbb{N}$. The authors also define a \textit{corresponding CFMM pool} of $\Phi$, denoted $\textit{CFMM}(\Phi)$. $\textit{CFMM}(\Phi)$ is the CFMM pool with reserves $(R_{x},R_{y})$ whose feasible set is described by pool invariant function $f()$ and pool constant $k=f(R_{x},R_{y})$. Conversely, $\Phi$ is the \textit{corresponding $\protocolName$ pool} of $\textit{CFMM}(\Phi).$ The authors note that $\textit{CFMM}(\Phi)$ changes every time the $\Phi$ pool reserves change. The protocol progresses in blocks, with one reserve update possible per block.
 
For an arbitrageur wishing to move the price of $\textit{CFMM}(\Phi)$ to $p$ from starting reserves $(R_{x,0},R_{y,0})$, let this require $\Delta_y>0$ tokens to be added to $\textit{CFMM}(\Phi)$, and $\Delta_x>0$ tokens to be removed from $\textit{CFMM}(\Phi)$. The same price in $\Phi$ is achieved by the following process:

\begin{enumerate}
    \item Adding $(1-\beta)\Delta_y$ tokens to $\Phi$ and removing $(1-\beta)\Delta_x$ tokens.
    \item Removing $\delta_x>0$ tokens such that:
    \begin{equation}
        P(R_{x,0}-(1-\beta)\Delta_x - \delta_x, R_{y,0}+(1-\beta) \Delta_y)=p.
    \end{equation}
    These $\delta_x$ tokens are added to the \textit{vault} of $\Phi$.
\end{enumerate} 

Vault tokens are periodically re-entered into $\Phi$ through what is effectively an auction process, where the tokens being re-added are in a ratio which approximates the external market price at the time. The main result of \cite{DiamondMcMenamin} is the proving that given a block producer interacts with $\Phi$ when the LVR parameter is $\beta$, and there is an LVR opportunity of $LVR$ in $CFMM(\Phi)$, the maximum LVR in $\Phi$ is $(1-\beta)LVR$. This results is stated formally therein as follows:

\begin{theorem}
    For a CFMM pool $CFMM(\Phi)$ with LVR of $L>0$, the LVR of $\Phi$, the corresponding pool in Diamond, has expectancy of at most $(1-\beta)L$.
\end{theorem}

In this paper we use the same base functionality of Diamond to restrict the LVR of block producers. Given a block producer wants to move the price of $CFMM(\Phi)$ to some price $p$ to extract maximal LVR $LVR$, the maximal LVR in $\Phi$ of $(1-\beta)LVR$ is also achieved by moving the price to $p$. An important point to note about applying LVR rebates as done in \cite{DiamondMcMenamin}, is that directly after tokens are placed in the vault, the pool constant drops. This must be considered when calculating the profitability of an arbitrageur extracting LVR from a Diamond pool. We do this when analyzing the profitability of $\protocolName$ in Section \ref{sec:properties}. Importantly, tokens are eventually re-added to the pool, and over time the expected value of the pool constant is increasing, as demonstrated in \cite{DiamondMcMenamin}. 

%% file: FormalProtocol.tex
\section{Our Protocol} \label{sec:Protocol}

We now outline the model in which we construct $\protocolName$, followed by a detailed description of $\protocolName$.

\subsection{Model}

In this paper we consider a blockchain in which all transactions are attempting to interact with a single $\protocolName$ pool between tokens $x$ and $y$. 

\begin{enumerate}
    \item A transaction submitted by a player for addition to the blockchain while observing blockchain height $\height$, is finalized in a block of height at most $\height+\revealTXDelay$, for some known $\revealTXDelay>0$.
    \item The token swap has an external market price $\MIFP$, which follows a Martingale process. 
    \item There exists a population of arbitrageurs able to frictionlessly trade at external market prices, who continuously monitor and interact with the blockchain.
    \item Encrypted orders are equally likely to buy or sell tokens at $\MIFP$, distributed symmetrically around $\MIFP$. 
\end{enumerate}

\subsection{Protocol Framework}

This section outlines the terminology and functionalities used in $\protocolName$. It is intended as a reference point to understand the core $\protocolName$ protocol. Specifically, we describe the possible transactions in $\protocolName$, the possible states that $\protocolName$ orders/order commitments can be in, and the possible actions of block producers. 
As in the protocol of Section \ref{sec:Diamond}, a $\protocolName$ pool $\Phi$ with reserves $(R_{x},R_{y})$ is defined with respect to a CFMM pool, denoted $CFMM(\Phi)$, with reserves $(R_{x},R_{y})$, a pricing function $P()$ under the restrictions of Section \ref{sec:CFMMs}, and a pool invariant function $f()$.

\subsubsection{Allocation Pools.}\label{sec:allocationPool}

Orders in $\protocolName$ are intended to interact with the AMM pool with some delay due to the commit-reveal nature of the orders. Therefore, we need to introduce the concept of allocated funds to be used when orders eventually get revealed. To do this, we define an \textit{allocation pool}.  For orders of size either $size_x$ or $size_y$ known to be of maximum size $max_x$ or $max_y$, the allocation pool consists of $(\lambda_x, \lambda_y)$ tokens such that:
\begin{equation}
    f(R_x,R_y)=f(R_x+max_x,R_y-\lambda_y)=f(R_x-\lambda_x,R_y+max_y).
\end{equation}
For such an allocation pool, let the total user tokens being sold be $\delta_x$ and $\delta_y$, with $\delta_x>\delta_y P(R_{x},R_{y})$. That is, there are more token $x$ being sold by users than the token $y$ required to match user orders against each other at the pool price  $P(R_{x},R_{y})$, causing an imbalance. This requires some additional $\Delta_y$ tokens from the allocation pool to satisfy the imbalance. If these orders are market orders\footnote{We omit a description of how to batch execute limit orders against allocation pools, leaving it as an implementation exercise. As long as limit orders follow the same size restrictions as specified in this paper, the properties of $\protocolName$ outlined in Section \ref{sec:properties} should not change.}, the execution price $p_e$ of these orders is such that $(\delta_y+\Delta_y)p_e=\delta_x$, and must satisfy:
\begin{equation}
    f(R_x,R_y)=f(R_x + (\delta_x-\delta_y p_e),R_y-\Delta_y).
\end{equation}
With these two restrictions, we can solve for $\Delta_y$ and $p_e$ given our specific pool pricing and invariant functions.\footnote{If $\delta_x<\delta_y P(R_{x},R_{y})$, we must remove $\Delta_x$ tokens from the allocation pool with $\delta_y p_e=\delta_x+\Delta_x$ satisfying $f(R_x,R_y)=f(R_x - \Delta_x,R_y+ (\delta_y-\frac{\delta_x}{p_e}))$} An example of batch settlement against an allocation pool with a Uniswap V2 pool as the corresponding CFMM pool is provided at the end of Section \ref{sec:Protocol}.

These restrictions for calculating the execution price and tokens to be removed from the allocation pool are not defined with respect to the tokens in the allocation pool. However, by definition of the allocation pool reserves, there are sufficient tokens in the allocation pool to handle any allowable imbalance (anything up to $max_x$ or $max_y$).

\subsubsection{Transaction Specifications.}

There are three types of transaction in our protocol.  To define these transactions, we need an \textit{LVR rebate function} $\beta: [0,1,...,Z, $ $ Z+1] \rightarrow [0,1]$. It suffices to consider $\beta()$ as a strictly decreasing function with $\beta(z)=0 \ \forall z\geq Z$.
\begin{enumerate}
    \item \textbf{Order}. These are straightforward buy or sell orders indicating a limit price\footnotemark[5], size and direction to be traded. Without loss of generality, we assume all orders in our system are executable. 
    
    \item \textbf{Order commitment transaction (OCT)}. These are encrypted orders known to be collateralized by either $max_x$ or $max_y$ tokens. The exact size, direction, price, and sender of an OCT sent from player $\player_i$ is hidden from all other players. This is possible using anonymous ZK proofs of collateral such as used in  \cite{FairTraDEXMcMenamin,TornadoCashWebsite,DPACCsMcMenamin}), which can be implemented on a blockchain in conjunction with a user-lead commit-reveal protocol, delay encryption scheme \cite{DelayEncryptionBurdges,FairPoSChiang} or threshold encryption scheme \cite{HelixAsayag,Penumbra}. An OCT must be inserted into the blockchain before that same OCT can be allocated liquidity in $\protocolName$.
    
    \item \textbf{Update transaction}. These transactions are executed in a block before any OCT is allowed to interact with the protocol pool (see Figure \ref{fig:V0LVERFlow}). Let the current block height be $H$. Update transactions take as input an \textit{allocation block height} $H_a\leq H$,  and pool price $p$. 
    Given an allocation block height of $H'_a$ in the previous update transaction, valid update transactions require $H_a>H'_a$. All of the inserted OCTs in blocks $[H'_a+1,...,H_a]$ are then considered as allocated. For any update transaction, we denote by $T_a\in \mathbb{Z}_{\geq 0}$ the number of OCTs being allocated.
    
    Given inputs $(H_a, p)$, the block producer moves the price of the pool to $p$. The producer receives $(1- \beta(H-H_a))$ of the implied change in reserves from this price move, as is done in \cite{DiamondMcMenamin}. 
    
    The producer must then deposit ($T_a \beta(H-H_a)max_y p, \ T_a \beta(H-H_a)\frac{max_x}{p})$ to an \textit{allocation pool} denoted $\Phi_{H_a}$, with ($T_a (1-\beta(H-H_a)) max_y p, \ T_a (1-\beta(H-H_a)) \frac{max_x}{p})$ being added to $\Phi_{H_a}$ from the AMM reserves. As such, the allocation pool contains ($T_a max_y p, \ T_a \frac{max_x}{p})$ tokens in total.
\end{enumerate}

In other words, if an allocation pool requires up to ($T_a  max_y p, T_a \frac{max_x}{p})$ tokens to trade with orders corresponding to the $T_a$ allocated OCTs, the block producer is forced to provide $\beta(H-H_a)$ of the tokens in the pool, with starting bid and offer prices equal to the pool price set by the block producer. This is used to incentivize the block producer to always choose a pool price equal to the external market price.

\subsubsection{Block Producer Action Set.}

Every block, a block producer has four possible actions to perform on OCTs and their orders. Orders in our system are batch-settled with other orders allocated at the same time, and against the liquidity in the respective allocation pool. 

\begin{enumerate}
    \item Insert OCTs to the blockchain.
    \item Allocate inserted OCTs. For a block producer adding a block at height $H$ to allocate any number (including 0) inserted OCTs with inserted height of at most $H_i$, the block producer must:
    \begin{enumerate}
        \item Submit an update transaction with inputs $(H_a=H_i, p)$, for some $p>0$.
        \item Allocate all unallocated OCTs with inserted height less than or equal to $H_i$. 
    \end{enumerate}
    \item Reveal allocated order. When a decrypted order corresponding to an OCT at height $H_a$ is finalized on the blockchain within $\revealTXDelay$ blocks after the corresponding OCT is allocated, it is considered revealed. 
    \item Execute revealed orders. $\revealTXDelay$ blocks after OCTs are allocated, any corresponding revealed orders are executed at a single clearing price for orders allocated at the same time. The final tokens in the allocation pool are redistributed proportionally to the allocating block producer and $\protocolName$ reserves. 
\end{enumerate}

\subsection{Protocol Outline}

Our protocol can be considered as two sub-protocols, a \textit{base protocol} proceeding in rounds corresponding to blocks in the blockchain (see Figure \ref{fig:V0LVERFlow}), and an \textit{allocation protocol} (Figure \ref{fig:allocationFlow}). As the blockchain progresses through the base protocol, at all heights $H>0$, the block producers has two key choices. The first is how many OCTs in the mempool to insert into the blockchain.  The second is whether or not to send an update transaction. 

There are two scenarios for an update transaction with inputs  $(H_a, p)$ and block height of the previous update transaction $H'_a$. Either $T_a=0$ or $T_a>0$. 
If $T_a=0$. the update transaction is equivalent to an arbitrageur operation on a Diamond pool with LVR-rebate parameter of $\beta(H-H_a)$ (see Section \ref{sec:Diamond}). 
If $T_a>0$, the arbitrageur must also deposit ($T_a  \beta(H-H_a))max_y p, T_a  \beta(H-H_a)\frac{max_x}{p})$ to the $H_a$ allocation pool $\Phi_{H_a}$, with ($T_a (1- \beta(H-H_a)) max_y p, T_a (1- \beta(H-H_a) \frac{max_x}{p})$ being added to $\Phi_{H_a}$ from the AMM reserves. 

After an allocation pool is created for allocated OCTs $\{oct_1,...,oct_{T_a}\}$, the orders corresponding to $\{oct_1,...,oct_{T_a}\}$ can be revealed for up to $\revealTXDelay$ blocks. This is sufficient time for any user whose OCT is contained in that set to reveal the order corresponding to the OCT. To enforce revelation, tokens corresponding to unrevealed orders are burned. After all orders have been revealed, or $\revealTXDelay$ blocks have passed, any block producer can execute revealed orders against the allocation pool at a clearing price which maximizes volume traded. Specifically, given an array of orders ordered by price, a basic smart-contract can verify that a proposed clearing price maximizes volume traded, as is done in \cite{FairTraDEXMcMenamin}. 

The final tokens in the allocation pool are redistributed to the allocating block producer and $\protocolName$ reserves. Adding these tokens directly to the pool (and not the vault as in the protocol from Section \ref{sec:Diamond}) allows the pool to update its price to reflect the information of the revealed orders.

\subsubsection{Example: Executing Orders Against the Allocation Pool.}\label{sec:Example}

This example details how one would batch execute orders against an allocation pool $\Phi_{H_a}$ replicating liquidity in a corresponding constant function MM, CFMM($\Phi$).
 Let the total tokens in the $\protocolName$ pool $\Phi$ before allocation be $(R_x,R_y)$, with CFMM($\Phi$) the Uniswap V2 pool. As such, $P(R_x,R_y)=\frac{R_x}{R_y}=p_0$.  Let the allocated OCTs be selling $\delta_x$ and $\delta_y$ tokens, with $\delta_y p_0 < \delta_x$. That is, there is an imbalance of tokens at $p_0$, with more token $x$ being sold than token $y$ at the price $p_0$. We will now derive the execution price $p_e$ for these orders.  

Given $\delta_y p_0 < \delta_x$, this means some $\Delta_y$ tokens from the allocation pool are required to fill the imbalance. Firstly, given the execution price is $p_e$, we know $(\delta_y+\Delta_y)p_e=\delta_x$. That is, the execution price equals $\frac{\delta_x}{\delta_y+\Delta_y}$. Secondly, the amount of $x$ tokens added to the allocation pool is $\delta_x-\delta_y p_e$. As the allocation pool provides liquidity equivalent to batch executing the orders against the CFMM, this means the pool invariant function would remain constant if those tokens were traded directly with CFMM($\Phi$). Specifically:
\begin{equation}
    R_x R_y=(R_x + (\delta_x-\delta_y p_e))(R_y-\Delta_y).
\end{equation}
From our first observation, we know $\Delta_y=\frac{\delta_x}{p_e}-\delta_y$, which we can rewrite as $\frac{1}{p_e}(\delta_x-\delta_y p_e)$. This gives:
\begin{equation}
    R_x R_y=R_x R_y +R_y(\delta_x-\delta_y p_e)-R_x\frac{1}{p_e}(\delta_x-\delta_y p_e)-\frac{1}{p_e}(\delta_x-\delta_y p_e)^2.
\end{equation}
Cancelling the first term on both sides, and dividing by $(\delta_x-\delta_y p_e)>0$ gives:
\begin{equation}
    0=R_y-R_x\frac{1}{p_e}-\frac{1}{p_e}(\delta_x-\delta_y p_e).
\end{equation}
Isolating $p_e$, we get:
\begin{equation}
    p_e=\frac{R_x+\delta_x}{R_y+\delta_y}.
\end{equation}

\input{Flow}

%% file: Flow.tex
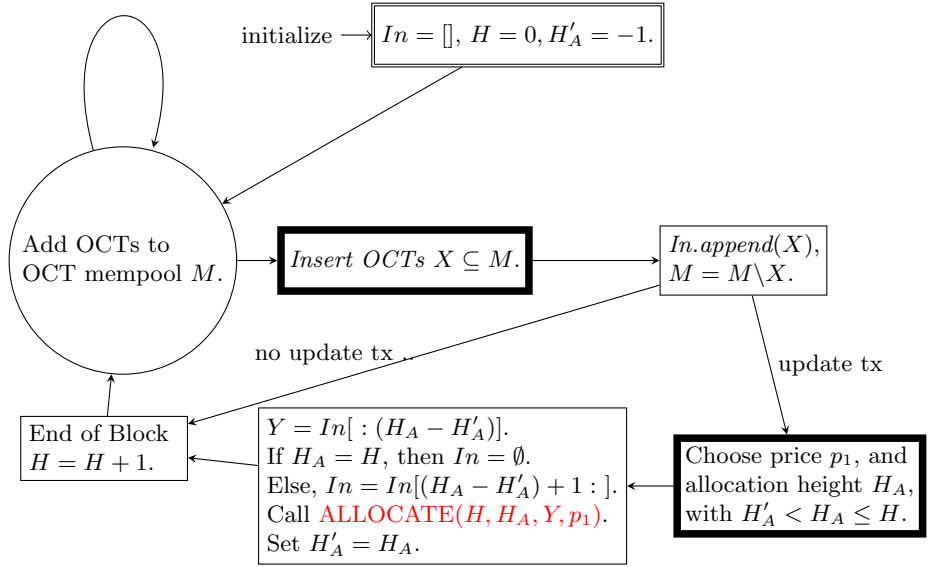
\begin{figure}[ht] 
\centering 
\begin{tikzpicture}[initial text = initialize]
    \node[state, initial, rectangle, accepting] (1) at (-.25,-2) {$In=[],$ $ H=0, H'_A=-1$.};
    \node[state, circle,align=left] (2) at (-5.5,-5) {Add OCTs to\\OCT mempool $M$.};
    \node[state, rectangle, line width=0.1cm] (6) at (-1.75,-5) {$\textit{Insert OCTs }X\subseteq M$.};
    \node[state, rectangle, text width=2cm] (3) at (2.75,-5) {$\textit{In.append}(X)$, $M=M \backslash X$.};
    \node[state, rectangle, line width=0.1cm,align=left] (4) at (3.5,-8) {Choose price $p_1$, and\\  allocation height $H_A$,\\ with $H'
    _A<H_A\leq H$.};
    \node[state, rectangle, align=left] (7) at (-1.25,-8) {$Y=In[ \ :(H_A- H'_A)]$.\\If $H_A=H$, then $In= \emptyset$.\\Else,
    $In=In[(H_A- H'_A)+1: \ ]$.\\ Call \textcolor{red}{ALLOCATE($H,H_A,Y,p_1$)}.\\Set $H'_A = H_A$.};
    \node[state,rectangle,text width=2cm] (5) at (-5.75,-7.5) {End of Block 
    $H=H+1$.};
    
    \draw [->,>=stealth] (1) edge[left] node{$ $} (2)
        (2) edge[above] node{$ $} (6)
        (6) edge[above] node{$ $} (3)
        (2) edge[loop above] node{$ $} (2)
        (3) edge[right] node{update tx} (4)
        (4) edge[above] node{$ $} (7)
        (7) edge[above] node{$ $} (5)
        (3) edge[left] node{{no update tx  ..}} (5)
        (5) edge[right] node{$ $} (2);
\end{tikzpicture}

\caption{Flow of V0LVER protocol, excluding the allocation protocol (see Figure \ref{fig:allocationFlow} for the allocation protocol). The double-border rectangle is the initialization state, thin single-border rectangles are state updates on-chain, while thick-bordered rectangles are block producer decisions/computations off-chain. The circle state is controlled by the network. Note that $In$, the array of inserted but unallocated OCTs, is an ordered array of sets of OCTs. For $1< a\leq len(In)$, $In[:a]$ returns an ordered sub-array of $In$ elements at indices $[1,...,a]$, while $In[a:]$ returns an ordered sub-array of $In$ elements at indices $[a,...,len(In)]$.}
\label{fig:V0LVERFlow}
\end{figure}

\begin{figure}[ht] 
\centering 
\begin{tikzpicture}[initial text = \textcolor{red}{ALLOCATE($H,H_A,Y,p_1$)}]
    \node[state, initial, rectangle, accepting,align=left] (1) at (3,-1.) {Let $T_A= \sum_{X\in Y} |X|$,\\max order sizes $max_x, \ max_y$.};
    \node[state, rectangle,align=left] (2) at (-1,-3) {Let LVR of CFMM($\Phi$) at $p_1$ be $L$.\\Give $(1-\beta(H-H_A))L$ to block producer.\\ Move $\Phi$ price to $p_1$.};
    \node[state, rectangle, align=left] (3) at (5.5,-3) {Create allocation pool $\Phi_{H_A}$,\\ with reserves $(T_A. max_y p_1, T_A \frac{max_x}{p_1})$.\\ Take $(1-\beta(H-H_A)) $ from $ \phi$,\\ and $\beta(H-H_A)$ from block producer. 
    };
    \node[state, rectangle] (4) at (2,-4.5) {\textbf{Reveal Orders}};
    \node[state, rectangle,align=left,line width = 0.1cm] (5) at (-1.75,-6) {Select clearing price $p_c$, with\\  $p_c$ maximizing volume of \\revealed $Y$ orders against $\Phi_{H_A}$.};
    \node[state, rectangle, align=left] (6) at (4.75,-6) {Verify $p_c$.\\
    Execute revealed $Y$ orders against $\Phi_{H_A}$ at $p_c$.\\
    Return $\Phi_{H_A}$ funds to $ \Phi$, block producer,\\ in ratio $1-\beta(H-H_A):\beta(H-H_A)$.};

    \draw [->,>=stealth] (1) edge[left] node{$ $} (2)
        (2) edge[above] node{$ $} (3)
        (3) edge[above] node{$ $} (4)
        (4) edge[left] node{$ $} (5)
        (5) edge[above] node{$ $} (6);
\end{tikzpicture}

\caption{Flow of allocation protocol for V0LVER pool $\phi$, initialized every time the ALLOCATE() function is called in Figure $\ref{fig:V0LVERFlow}$. The Reveal Orders state happens by some block after height $H$. As in the previous figure, the double-border rectangle is the initialization state, thin single-border rectangles are state updates on-chain, while thick-bordered rectangles are block producer decisions/computations off-chain.}
\label{fig:allocationFlow}
\end{figure}
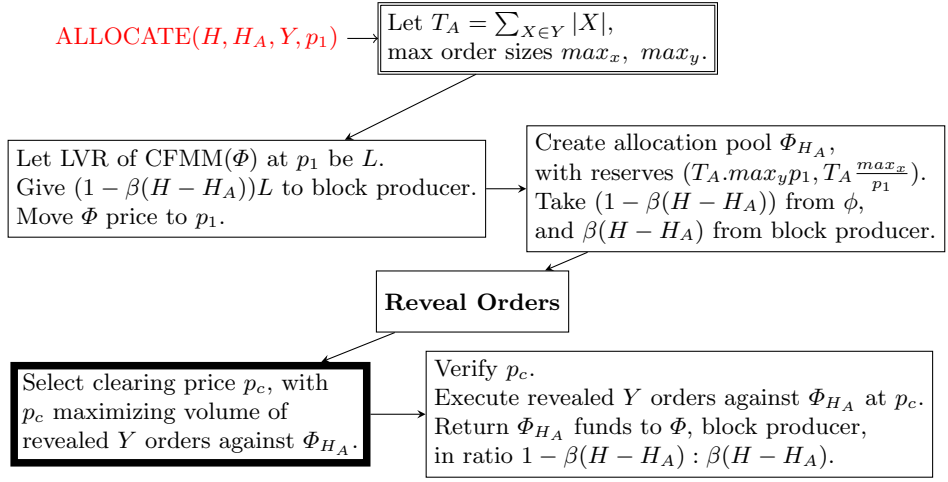

%% file: ProtocolProperites.tex
\section{Protocol Properties}\label{sec:properties}

The goal of this section is to show that the expected execution price of any user order is the external market price when the order is allocated, excluding at most impact and fees. Firstly, note that an update transaction prior to allocation moves the pool reserves of a $\protocolName$ pool identically to an LVR arbitrage transaction in Section \ref{sec:Diamond}.
If $T_a=0$, from \cite{DiamondMcMenamin} we know the block producer moves the pool price to the max LVR price which is the external market price, and the result follows trivially.

Now instead, assume $T_a>0$. Let the reserves of a $\protocolName$ pool $\Phi$ before the update transaction be $(R_{x,0},R_{y,0})$. Given an external market price of $\MIFP$, from Section \ref{sec:CFMMs} we know the max LVR occurs by moving the pool reserves to some $(R_{x,m},R_{y,m})$ with $\frac{R_{x,m}}{R_{y,m}}=\MIFP$. Without loss of generality, let $\frac{R_{x,0}}{R_{y,0} }<\frac{R_{x,m}}{R_{y,m}}$. Let the block producer move the pool price to $p$ corresponding to reserves in the corresponding CFMM pool of $(R_{x,p},R_{y,p})$. For a non-zero $\beta()$, this means the tokens in $\Phi$ not in the vault (as per the protocol in Section \ref{sec:Diamond}) are $(R'_{x,p},R'_{y,p})=(b R_{x,p},b R_{y,p})$ for some $b<1$. This is because some tokens in $\Phi$ are removed from the pool and placed in the vault, while maintaining $\frac{R'_{x,p}}{R'_{y,p}}=p$. 

There are three payoffs of interest here. For these, recall that by definition of the external market price, the expected imbalance of an encrypted order in our system is 0 at the external market price.
\begin{enumerate}
    \item \textbf{Payoff of block producer vs. AMM pool}: $(1-\beta())(R_{x,0}-R_{x,p}+ (R_{y,0}-R_{y,p})\MIFP)$.
    
    \item \textbf{Payoff of block producer vs. users}: Against a block producer's own orders, the block producer has 0 expectancy. Against other player orders, the block producer strictly maximizes her own expectancy when $(R_{x,p},R_{y,p})=(R_{x,m},R_{y,m})$. Otherwise the block producer is offering below $\MIFP$ against expected buyers, or bidding above $\MIFP$ to expected sellers.

    \item \textbf{Payoff of users vs. AMM pool}: Consider a set of allocated orders executed against the allocation pool, corresponding to the pool receiving $\delta_x$ and paying $\delta_y$ tokens. By definition of the allocation pool, this $(\delta_x,\delta_y)$ is the same token vector that would be applied to the CFMM pool with reserves $(b R_{x,p},b R_{y,p})$ if those orders were batch executed directly against the CFMM. Let these new reserves be $(b R_{x,1},b R_{y,1})$. Thus the profit of these orders is $b(1-\beta())(R_{x,p}-R_{x,1}+ (R_{y,p}-R_{y,1})\MIFP)$.
\end{enumerate}

\subsubsection*{Optimal strategy for block producer}

Let the block producer account for $\alpha\in[0,1]$ of the orders executed against the allocation pool. The maximum payoff of the block producer against the AMM pool is the maximum of the sum of arbitrage profits (Payoff 1) and profits of block producer orders executed against the pool ($\alpha$ of Payoff 3). Thus, the utility function to be maximized is:
\begin{equation}
    (1-\beta())(R_{x,0}-R_{x,p}+ (R_{y,0}-R_{y,p})\MIFP) + \alpha \Big( b (1-\beta())(R_{x,p}-R_{x,1}+ (R_{y,p}-R_{y,1})\MIFP) \Big).
\end{equation}
This is equal to 
\begin{equation}
    (1-\alpha b)(1-\beta())\big(R_{x,0}-R_{x,p}+ (R_{y,0}-R_{y,p})\MIFP \big)+\alpha b (1-\beta())\big(R_{x,0}-R_{x,1}+ (R_{y,0}-R_{y,1})\MIFP \big).
\end{equation}

We know the second term is maximized for $(R_{x,1},R_{y,1})=(R_{x,m},R_{y,m})$, as this corresponds to LVR. Similarly, the first term is also maximized for $(R_{x,p},R_{y,p})=(R_{x,m},R_{y,m})$. Given $(R_{x,p},R_{y,p})=(R_{x,m},R_{y,m})$, block producers have negative expectancy for $\alpha>0$, as this reduces the probability that $(R_{x,1},R_{y,1})=(R_{x,m},R_{y,m})$ by increasing the likelihood of an imbalance at $p$. As such, block producers are strictly incentivized to set $p=\MIFP$, and not submit OCTs to the protocol ($\alpha=0$) for Payoffs 1 and 3. 
Now consider the payoff for the block producer against user orders (Payoff 2). We have already argued in the description of Payoff 2 that this is maximized with $(R_{x,p},R_{y,p})=(R_{x,m},R_{y,m})$. 

Therefore, moving the pool price $p$ to $\MIFP$ is a dominant strategy for the block producer.
Given this, we can see that the expected execution price for a client is $\MIFP$ excluding impact and fees, with impact decreasing in expectancy in the number of orders allocated. 
The payoff for the AMM against the block producer via the update transaction is $(1-\beta())LVR$, while the payoff against other orders is at least 0.

\subsection{Minimal LVR}

In the previous section, it is demonstrated that user-level MEV is prevented in $\protocolName$, with users trading at the external market price in expectancy, excluding fees. However, we have thus far only proved that LVR in a $\protocolName$ pool is $(1-\beta())$ of the corresponding CFMM pool. As in \cite{DiamondMcMenamin}, under competition among block producers, the LVR rebate function has a strong Nash equilibrium at $\beta(0)$, meaning LVR is also minimized.

To see this, we can use a backwards induction argument. Consider the first block producer allowed to send an update transaction with $\beta(H-H_a)=0$ for a block at height $H$ (meaning $H_a=H'_a+1$). This block producer can extract all of the LVR, and is required to provide no liquidity to the allocation pool. As LVR is arbitrage, all block producers do this. 

A block producer at height $H-1$ knows this. Furthermore, extracting $(1-\beta((H-1)-H_a))>0$ of the LVR has positive utility for all block producers, while trading with $\beta((H-1)-H_a)>0$ of allocated OCTs around the external market price also has a positive utility (Payoff 2 in Section \ref{sec:properties}). As such, sending an update transaction at height $H-1$ is dominant. Following this argumentation, a block producer at height $H-i\geq H_a$ always sends an update transaction as they know the block producer at height $(H+1)-i$ always sends an update transaction. This means the block producer at height $H'_a+1$ always sends an update transaction $\forall \ H'_a$, which corresponds to an LVR rebate function value of $\beta(0)$ in equilibrium.

In reality, frictionless arbitrage against the external market price in blockchain-based protocols is likely not possible, and so LVR extraction has some cost. As such, the expected value for $\beta()$ may be less than $\beta(0)$. Deploying $\protocolName$, and analyzing $\beta()$ across different token pairs, and under varying costs for block producers makes for interesting future work. 

%% file: Discussion.tex
\section{Discussion}

If a $\protocolName$ pool allows an OCT to be allocated with $\beta()=0$, $\protocolName$ effectively reverts to the corresponding CFMM pool, with MEV-proof batch settlement for all simultaneously allocated OCTs, albeit without LVR protection for the pool. To see this, note that as $\beta()=0$, the block producer can fully extract any existing LVR opportunity, without requiring a deposit to the allocation pool. As such, the expected price of the allocation pool is the external market price, with orders executed directly against the $\protocolName$ reserves at the external market price, excluding fees and impact. Importantly, there is never any way for the block producer to extract any value from allocated orders. This is because the settlement price for an OCT is effectively set when it allocated, before any price or directional information is revealed about the corresponding order.

Allocation of tokens to the allocation pool has an opportunity cost for both the $\protocolName$ pool and the block producer. Given the informational superiority of the block producer, allocating tokens from the pool requires the upfront payment of a fee to the pool. Doing this anonymously is important to avoid MEV-leakage to the block producer. One possibility is providing an on-chain verifiable proof of membership to set of players who have bought pool credits, where a valid proof releases tokens to cover specific fees, as in \cite{TornadoCashWebsite,FairTraDEXMcMenamin}. Another possibility is providing a proof to the block-producer that the user has funds to pay the fee, with the block-producer paying the fee on behalf of the user. A final option based on threshold encryption \cite{Penumbra} is creating a state directly after allocation before any more allocations are possible, in which allocated funds are either used or de-allocated. All of these proposals have merits and limitations, but further analysis of these are beyond the scope of this work.

%% file: Conclusion.tex
\section{Conclusion}

$\protocolName$ is an AMM based on an encrypted transaction mempool in which LVR and MEV are protected against. $\protocolName$ aligns the incentives of users, passive liquidity providers and block producers. This is done by ensuring the optimal block producer strategy under competition among block producers simultaneously minimizes LVR against passive liquidity providers and MEV against users. 

Interestingly, the exact strategy equilibria of $\protocolName$ depend on factors beyond instantaneous token maximization for block producers. This is due to risks associated with liquidity provision and arbitrage costs. On one hand, allocating OCTs after setting the pool price to the external market price, and providing some liquidity to OCTs around this price should be positive expectancy for block producers. Similarly, increasing the number of OCTs should also reduce the variance of block producer payoffs. On the other hand, there are caveats in which all OCTs are informed and uni-directional. Analyzing these trade-offs for various risk profiles and trading scenarios makes for further interesting future work.

%% file: main.bbl
\begin{thebibliography}{10}
\providecommand{\url}[1]{\texttt{#1}}
\providecommand{\urlprefix}{URL }
\providecommand{\doi}[1]{https://doi.org/#1}

\bibitem{Adams2020UniswapVC}
Adams, H., Zinsmeister, N., Robinson, D.: {Uniswap V2 Core} (2020),
  \url{https://uniswap.org/whitepaper.pdf}

\bibitem{HelixAsayag}
Asayag, A., Cohen, G., Grayevsky, I., Leshkowitz, M., Rottenstreich, O.,
  Tamari, R., Yakira, D.: {Helix: A Fair Blockchain Consensus Protocol
  Resistant to Ordering Manipulation}. IEEE Transactions on Network and Service
  Management  \textbf{18}(2),  1584--1597 (2021).
  \doi{10.1109/TNSM.2021.3052038}

\bibitem{ZCash}
Ben-Sasson, E., Chiesa, A., Garman, C., Green, M., Miers, I., Tromer, E.,
  Virza, M.: {Zerocash: Decentralized Anonymous Payments from Bitcoin}. In:
  2014 IEEE Symposium on Security and Privacy. pp. 459--474. IEEE Computer
  Society, New York, NY, USA (2014)

\bibitem{FrequentBatchAuctionsBudish}
Budish, E., Cramton, P., Shim, J.: { The High-Frequency Trading Arms Race:
  Frequent Batch Auctions as a Market Design Response *}. The Quarterly Journal
  of Economics  \textbf{130}(4),  1547--1621 (07 2015).
  \doi{10.1093/qje/qjv027}, \url{https://doi.org/10.1093/qje/qjv027}

\bibitem{DelayEncryptionBurdges}
Burdges, J., De~Feo, L.: Delay encryption. In: Canteaut, A., Standaert, F.X.
  (eds.) Advances in Cryptology -- EUROCRYPT 2021. pp. 302--326. Springer
  International Publishing, Cham (2021)

\bibitem{DEXCritiqueCapponi}
Capponi, A., Jia, R.: {The Adoption of Blockchain-based Decentralized
  Exchanges}. \url{https://arxiv.org/abs/2103.08842} (2021), accessed:
  10/02/2023

\bibitem{FairPoSChiang}
Chiang, J.H., David, B., Eyal, I., Gong, T.: Fairpos: Input fairness in
  proof-of-stake with adaptive security.
  \url{https://eprint.iacr.org/2022/1442} (2022), accessed: 23/01/2023

\bibitem{FlashBoys2.0}
Daian, P., Goldfeder, S., Kell, T., Li, Y., Zhao, X., Bentov, I., Breidenbach,
  L., Juels, A.: {Flash Boys 2.0: Frontrunning, Transaction Reordering, and
  Consensus Instability in Decentralized Exchanges}.
  \url{https://arxiv.org/abs/1904.05234} (2019), accessed: 19/01/2022

\bibitem{FlashbotsExploreWebsite}
Flashbots: \url{https://explore.flashbots.net}, accessed: 11/10/2022

\bibitem{DynamicAMMsKrishnamachari}
Krishnamachari, B., Feng, Q., Grippo, E.: {Dynamic Automated Market Makers for
  Decentralized Cryptocurrency Exchange}. In: 2021 IEEE International
  Conference on Blockchain and Cryptocurrency (ICBC). pp.~1--2 (2021).
  \doi{10.1109/ICBC51069.2021.9461100}

\bibitem{DPACCsMcMenamin}
McMenamin, C., Daza, V.: Dynamic, private, anonymous, collateralizable
  commitments vs. mev. \url{https://arxiv.org/abs/2301.12818} (2022).
  \doi{10.48550/ARXIV.2301.12818}, accessed: 31/01/2023

\bibitem{FairTraDEXMcMenamin}
McMenamin, C., Daza, V., Fitzi, M., O'Donoghue, P.: {FairTraDEX: A
  Decentralised Exchange Preventing Value Extraction}. In: Proceedings of the
  2022 ACM CCS Workshop on Decentralized Finance and Security. p. 39–46.
  DeFi'22, Association for Computing Machinery, New York, NY, USA (2022).
  \doi{10.1145/3560832.3563439}, \url{https://doi.org/10.1145/3560832.3563439}

\bibitem{DiamondMcMenamin}
McMenamin, C., Daza, V., Mazorra, B.: {Diamonds are Forever,
  Loss-Versus-Rebalancing is Not}. \url{https://arxiv.org/abs/2210.10601}
  (2022). \doi{10.48550/ARXIV.2210.10601}, accessed: 04/01/2023

\bibitem{LVRRoughgarden}
Milionis, J., Moallemi, C.C., Roughgarden, T., Zhang, A.L.: {Quantifying Loss
  in Automated Market Makers}. In: Zhang, F., McCorry, P. (eds.) Proceedings of
  the 2022 ACM CCS Workshop on Decentralized Finance and Security. ACM (2022)

\bibitem{Park2021TheCF}
Park, A.: {The Conceptual Flaws of Constant Product Automated Market Making}.
  ERN: Other Microeconomics: General Equilibrium \& Disequilibrium Models of
  Financial Markets  (2021)

\bibitem{Penumbra}
Penumbra: \url{https://penumbra.zone/}, accessed: 23/01/2023

\bibitem{QuantifyingExtractableValueGervais}
Qin, K., Zhou, L., Gervais, A.: {Quantifying Blockchain Extractable Value: How
  dark is the forest?} In: 2022 IEEE Symposium on Security and Privacy (SP).
  pp. 198--214 (2022). \doi{10.1109/SP46214.2022.9833734}

\bibitem{BatchCFMMsRamseyer}
Ramseyer, G., Goyal, M., Goel, A., Mazières, D.: Batch exchanges with constant
  function market makers: Axioms, equilibria, and computation.
  \url{https://arxiv.org/abs/2210.04929} (2022).
  \doi{10.48550/ARXIV.2210.04929}, accessed: 26/01/2023

\bibitem{duneToxicityQuery}
@thiccythot: \url{https://dune.com/thiccythot/uniswap-markouts}, accessed:
  10/02/2023

\bibitem{TornadoCashWebsite}
{Tornado Cash}: \url{https://tornadocash.eth.link/}, accessed: 31/01/2023

\end{thebibliography}
